\documentclass[%
 aip, apl,%
 amsmath,amssymb,
 reprint,%
unsortedaddress
]{revtex4-1}
\pdfoutput=1 

\usepackage{graphicx}
\usepackage{dcolumn}
\usepackage{bm}

\begin{document}


\title{Oblique propagation of electrons in crystals of germanium and silicon at sub-Kelvin temperature in low electric fields}



\author{B.~Cabrera}
\email[Email address: ]{cabrera@stanford.edu}

\author{M.~Pyle}

\author{R.~Moffatt}
\affiliation{Department of Physics, Stanford University, Stanford, CA 94305, USA}

\author{K.M.~Sundqvist}

\author{B.~Sadoulet}
\affiliation{Department of Physics, University of California at Berkeley, CA 94720, USA}


\date{\today}

\begin{abstract}
We show that oblique propagation of electrons in crystals of Ge and Si, where the electron velocity does not follow the electric field even on average, can be explained using standard anisotropic theory for indirect gap semiconductors.  These effects are pronounced at temperatures below $\sim$1~K and for electric fields below $\sim$5~V/cm because inter-valley transitions are energetically suppressed forcing electrons to remain in the same band valley throughout their motion and the valleys to separate in position space. To model, we start with an isotropic approximation which incorporates the average properties of the crystals with one phonon mode, and include the ellipsoidal electron valleys by transforming into a momentum space where constant energy surfaces are spheres.  We include comparisons of simulated versus measured drift velocities for holes and electrons, and explain the large discrepancy between electrons and holes for shared events in adjacent electrodes.  
\end{abstract}


\maketitle




When low energy recoils occur in cryogenic Ge or Si crystal radiation detectors, such as those used for the CDMS (Cryogenic Dark Matter Search) experiments,~\cite{Akerib:2005} energy is contained within a few hundred $\mu$m as a cloud of energetic electrons rapidly relax to very low energy initially by exciting more electrons and then by emitting optical phonons.  This cloud has an equal number of electrons and holes which begin to separate and move in opposite directions in the applied electric field while emitting low energy acoustic phonons.~\cite{Luke:1988}
In the CDMS Ge detectors electrons behave very differently from holes. For example, with inner and guard ring charge electrodes we find that when electrons are pulled towards the electrodes the fraction of shared events is ten times greater than when holes are pulled towards the same electrodes, suggesting ten times larger lateral straggle for electrons.~\cite{Cabrera:2010}

The existence of these anisotropic effects have been known since the 1950s, but in classical calculations~\cite{Jacoboni:1983, Ridley:1982} only approximations of this physics could be incorporated.  The computing power available today allows more rigorous computation taking into account energy momentum conservation, the full shape and anisotropy of the carrier and phonon dispersion relations, and the various scattering processes (e.g. intra-valley and  inter-valley).~\cite{Aubry:2009,Sundqvist:2009}

However, it is important to develop simpler models to gain physical insight of the important processes and to validate the more ambitious transport calculations. Such simplifications may also speed up Monte Carlo detector simulations. In this spirit, we present a simple model where for holes the anisotropic nature of the crystal is taken into account by considering an isotropic medium with the average properties of the anisotropic crystals,~\cite{Ridley:1982} and additionally for electrons we transform the anisotropic momentum space to an isotropic space.~\cite{Herring:1956}
We show that this model successfully reproduces the drift velocities of the holes and the electrons using one adjustable parameter (deformation potential $\Xi$) for each.  Both Ge and Si are indirect gap semiconductors with electron valleys away from the momentum space origin, and the hole valence band with its minimum at the origin of momentum space. For holes, the model assumes a $\Gamma$ spherical valley, and for electrons the model includes the four $L$ elliptical valleys along [111] axes for Ge or the six $\Delta$ valleys along [100] axes for Si by transforming each into a momentum space where the constant energy elliptical surfaces become spheres.~\cite{Herring:1956, Ridley:1982}  Then transforming back to real space we obtain the lowest order anisotropic behavior for the electrons. These anisotropic effects~\cite{Sasaki:1958} dominate at temperatures below $\sim$1~K and for electric fields below $\sim$5~V/cm because inter-valley transitions are energetically suppressed forcing the electrons to remain in the same valley throughout their motion.  

Starting with an isotropic model for holes, we consider the case of an incident hole with energy $\epsilon_{\vec k}$ and wave vector $\vec k$ scattering off of the lattice and emitting a phonon of energy $\hbar \omega$ and momentum $\vec q$, and the hole has final state energy $\epsilon_{\vec k'}$ and wave vector $\vec k'$. Below 1~K thermal phonons are suppressed so that only spontaneous phonon emission occurs. Energy and momentum conservation requires $\epsilon_{\vec k} - \epsilon_{\vec k'} = \hbar \omega$ and $\vec k - \vec k' = \vec q + \vec K$ where the reciprocal lattice vector $\vec K=0$ at our low temperature and low electric field.  Since only longitudinal phonon modes couple in an isotropic medium, we assume one mode with $\omega / q = s_L$ and isotropic speed of sound $s_L$. We assume a parabolic band minimum, so that $\epsilon_{\vec k} = {\hbar^2k^2}/{(2m_h)}$ where $m_h$ is the hole effective mass. In Table~\ref{tab:constants}, we collect the constants needed for these calculations.     

\begin{table}[ht]
  \begin{center}
 \caption{Physical constants for Si and Ge crystals. The isotropic hole effective mass $m_h$, and the anisotropic electron effective masses $m_\parallel$ and $m_\perp$ are $\parallel$ and $\perp$, respectively, to the conduction valley axes, and conductivity effective mass $3/m_c = 1/m_\parallel + 2/m_\perp$. The incident energy per final electron-hole pair is $\epsilon_{eh}$, $s_L$ the speed of sound, 
and $l_0 = {\pi \hbar^4 \rho}/{(2 m^3 \Xi^2)}$ is the characteristic range for carrier scattering where $\Xi_1$ (from~\cite{Jacoboni:1983}) or $\Xi_{\text fit}$ (fit to data~\cite{Sundqvist:2009}) is the deformation potential.}
  \label{tab:constants}
  \vspace{.1in}
  \begin{tabular}{|c|c|c|c|c|}
    \hline
    &\multicolumn{2}{|c|}{Silicon}& \multicolumn{2}{|c|}{Germanium}\\\hline
    & Electrons & Holes & Electrons & Holes \\\hline
    $m_h/m_e$           &  -    & 0.5   &   -   & 0.35  \\\hline
    $m_{\parallel}/m_e$ & 0.91 &  -   & 1.58  &  -  \\\hline
    $m_{\perp}/m_e$     & 0.19  &  -  & 0.081 &  -  \\\hline
    $m_c/m_e$           & 0.26  &   -   & 0.12  &   -   \\\hline
    $\epsilon_{eh}$ (eV)      & \multicolumn{2}{|c|}{3.84} & \multicolumn{2}{|c|}{3.00} \\\hline
    $s_L$ (km/s)      & \multicolumn{2}{|c|}{9.0} & \multicolumn{2}{|c|}{5.4} \\\hline
    $\rho$ (g/cm$^3$)  & \multicolumn{2}{|c|}{2.335} & \multicolumn{2}{|c|}{5.323} \\\hline
    $\Xi_1$ (eV)        & 9.0  & 5.0  & 11.0  & 4.6  \\\hline
    $\Xi_{\text fit}$ (eV)        &  -  &  -   & 11.0  & 3.4  \\\hline
    $l_0$ ($\mu$m)        & 16.9  &  7.5  & 257  & 108 \\\hline
  \end{tabular}
  \end{center}
 \end{table}

From momentum conservation and the law of cosines $k'^2=k^2+q^2-2kq\cos\theta$ with $\theta$ the angle between $\vec k$ and $\vec q$, and from energy conservation $k^2 - k'^2 = ({2 m_h}/{\hbar^2}) \hbar \omega = ({2 m_h s}/{\hbar}) q=2 k_s q$ where $k_s = {m_h s}/{\hbar}$.  Combining we obtain $q=2(k \cos\theta-k_s)$, which has solutions only if $k>k_s$ or the hole velocity $v$ is greater than the speed of sound $s$, analogous to Cherenkov radiation of photons.

For simulations,~\cite{Jacoboni:1983} we use Fermi's golden rule to compute the azimuthally symmetric differential phonon emission rate $P(k,\theta) \sin\theta d\theta$ for angles between $\theta$ and $\theta +d\theta$ relative to the incident hole direction
\begin{equation}
P(k,\theta) \sin\theta d\theta =\frac{s}{l_0} \left(\frac {k}{k_s}\right)^2 \left(\cos\theta - \frac{k_s}{k}\right)^2 \sin\theta d\theta
\label{equ:1}
\end{equation}
where $0 \leq \theta \leq \cos^{-1}(k_s / k) < {\pi}/{2}$, and $l_0 = {\pi \hbar^4 \rho}/{(2 m_h^3 \Xi^2)}$ is the characteristic scattering range~\cite{Ridley:1982}  with deformation potential $\Xi$.
Integrating over $\theta$ we obtain the total isotropic hole-phonon scattering rate
\begin{equation}
 \frac{1}{\tau_{hp}}=\frac{s}{3 l_0} \left(\frac {k}{k_s}\right)^2 \left(1 - \frac{k_s}{k}\right)^3\text{ for } k \geq k_s
\label{equ:2}
\end{equation}
Finally, we compute the hole scattering angle and obtain
\begin{equation}
\cos\phi=\frac{k^2-2k_s(k\cos\theta-k_s)-2(k\cos\theta-k_s)^2}{k\sqrt{k^2-4k_s(k\cos\theta-k_s)}}
\label{equ:3}
\end{equation}
where the angle ranges from $0 \leq \phi \leq \pi$ for $k>2k_s$ and from $0 \leq \phi \leq \pi/2$ for $k_s<k<2k_s$.

\begin{figure}[h]
  \begin{center}
  \includegraphics[width=3.2in]{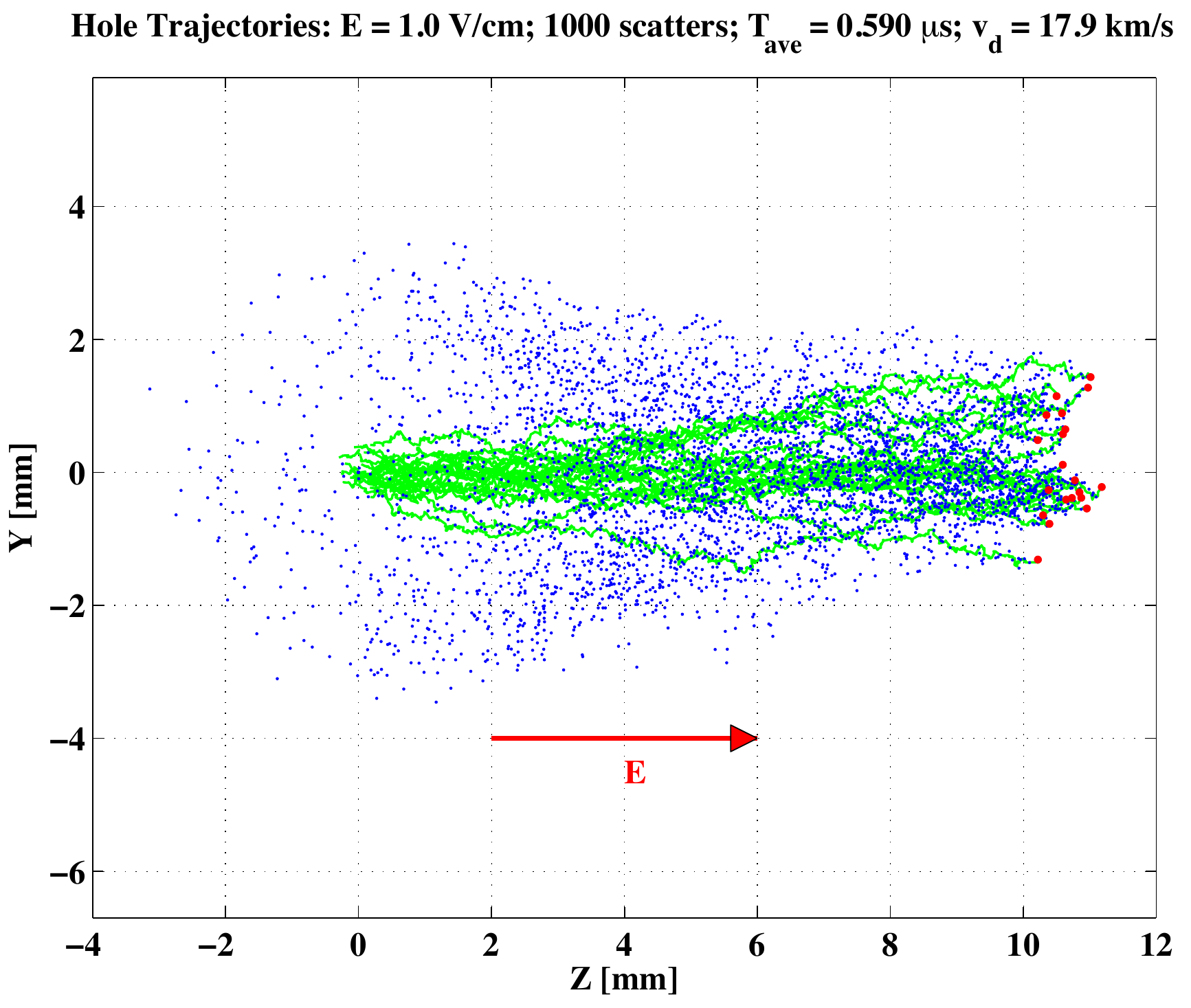} 
  \vspace{-0.1in}
  \end{center}
  \caption{Simulation of hole propagation through Ge cyrstal at zero temperature with a 1 V/cm applied electric field.  Shown are the hole trajectories (green), the emitted phonon positions (blue), and the final hole positions (red).  On average 1000 scatters took 0.59 $\mu$s, traversed 10.8 mm and transverse straggle standard deviation of 0.71 mm.}
  \label{fig:hole_sim}
\end{figure}

\begin{figure}[h]
  \begin{center}
  \includegraphics[width=3.2in]{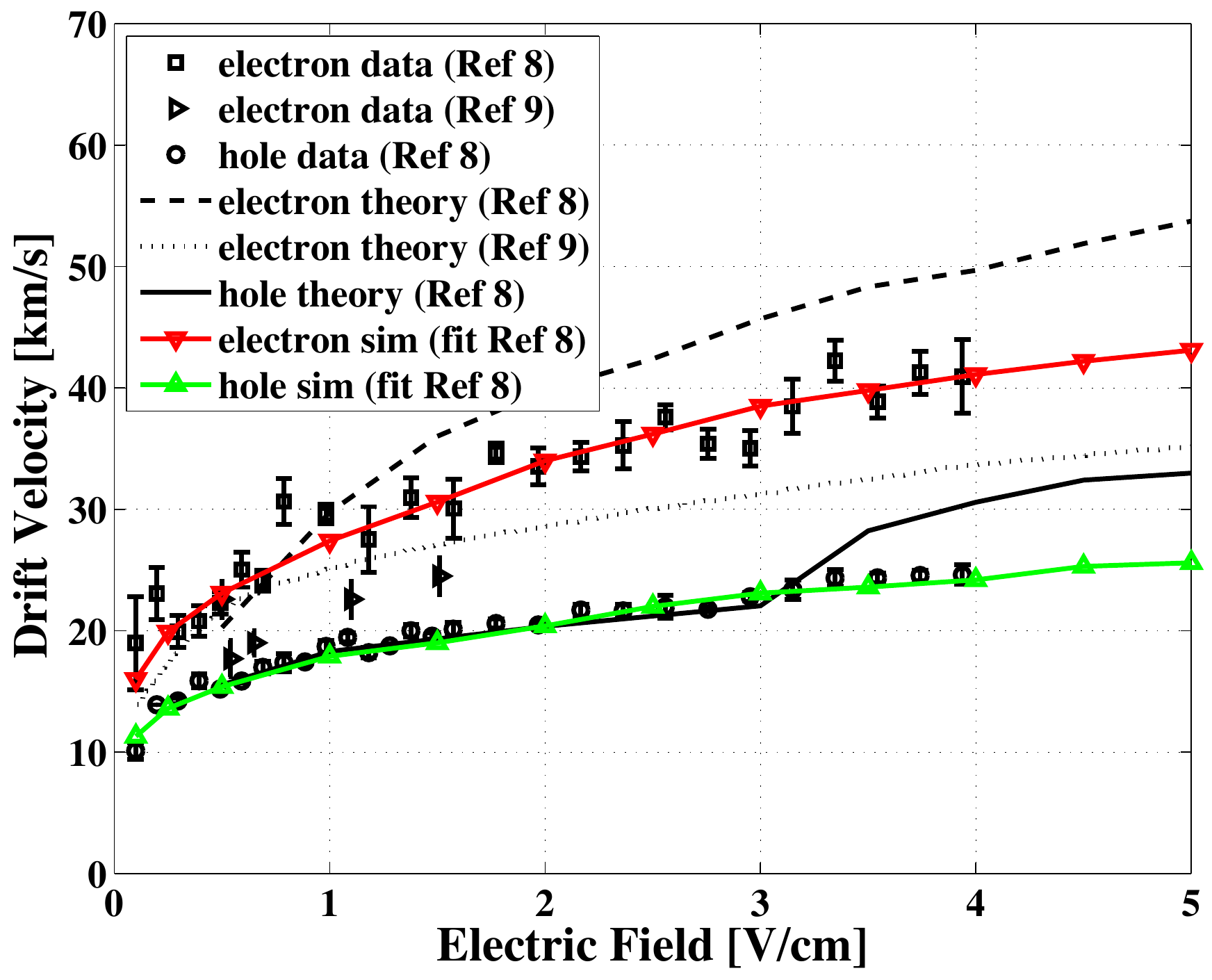} 
  \vspace{-0.1in}
  \end{center}
  \caption{ Comparison of the measured and simulated drift velocities as a function of electric field along [100] axis for ultrapure ($N_a = 1.83 \times 10^{11}$ cm$^{-3}$) Ge crystals at 31 mK.~\cite{Sundqvist:2009}  One parameter fits to Ref 8 data were obtained by setting the deformation potential for holes to $\Xi_{fit} = 3.4$ eV (green) and for electrons to $\Xi_{fit} = 11.0$ eV (red). Also shown are published theory curves for electrons~\cite{Sundqvist:2009,Aubry:2009} and for holes.~\cite{Sundqvist:2009}  }
  \label{fig:drift_compare}
\end{figure}

Using these equations for hole propagation in the isotropic medium, we have performed simulations at various applied voltages.   In Fig.~\ref{fig:hole_sim} we show 20 hole trajectories each with 1000 scattering events. The spatial dispersion about the mean displacement is a diffusive process, and increases with the square root of distance or time. In Table~\ref{tab:parameters} we show analytic versions of many of the parameters of interest in the limit $k >> k_s$ , including the lateral dispersion, average phonon energy, average hole energy, and hole drift velocity. Finally, we compare with the Ge drift velocity measurements made recently with CDMS detectors,~\cite{Sundqvist:2009} and we fit the simulation to the data with $\Xi_{\text fit} = 3.4$ eV. Note from Fig.~\ref{fig:drift_compare} that the shape of the hole drift velocity versus electric field is reproduced.

\begin{table}[ht]
  \caption{Drift velocity parameters for Si and Ge, in terms of a unit-less $\alpha(E) = (e E l_0)/(2 \epsilon_s)$ which is proportional to the applied electric field $E$ and $\epsilon_s = \frac{1}{2}m_c s^2$.  For isotropic models, in the limit of $k >> k_s$, we compute the drift velocity $v_d$, average carrier velocity $\bar v$, average phonon energy  $h \bar \omega$, average carrier energy $\bar \epsilon$, mean free path $\lambda$ and component parallel to field, and standard deviation of straggle $\sigma_\perp $. The values in last four columns are for $E = $ 1 V/cm and $x = $ 1 cm.}
  \label{tab:parameters}
  \begin{center}
  \begin{tabular}{|l|c|c|c|c|c|}
    \hline
    \multicolumn{2}{|c|}{  }&\multicolumn{2}{|c|}{Silicon}& \multicolumn{2}{|c|}{Germanium}\\\hline
   & units & Electrons & Holes & Electrons & Holes \\\hline
    $v_d = 1.31 s \alpha^{1/5}$ & km/s & 20.1 & 15.0 & 29.7 & 20.2 \\\hline
    $\bar v = 1.38 s \alpha^{2/5}$ & km/s & 35.9 & 19.9 & 131 & 60.3 \\\hline
    $\hbar \bar \omega = 4.14 \epsilon_s \alpha^{2/5}$ & meV & 0.71 & 0.77 & 0.72 & 0.97 \\\hline
    $\bar \epsilon = 1.90 \epsilon_s \alpha^{4/5}$ & meV & 0.95 & 0.56 & 5.8 & 3.6 \\\hline
    $\lambda = 2.17 l_0 \alpha^{-2/5}$ & $\mu$m & 12.7 & 10.2 & 31.8 & 29.1 \\\hline
    $\lambda_{\parallel} = 2.07 l_0 \alpha^{-3/5}$ & $\mu$m & 7.1 & 7.7 & 7.2 & 9.7 \\\hline
    $\sigma_\perp = 1.70 \sqrt{x l_0} \alpha^{-1/10}$ & mm & 0.54 & 0.42 & 1.33 & 1.05 \\\hline
  \end{tabular}
  \end{center}
\end{table}

\begin{figure}[ht]
\vspace{0.1in}
  \begin{center}
  \includegraphics[width=3.2in]{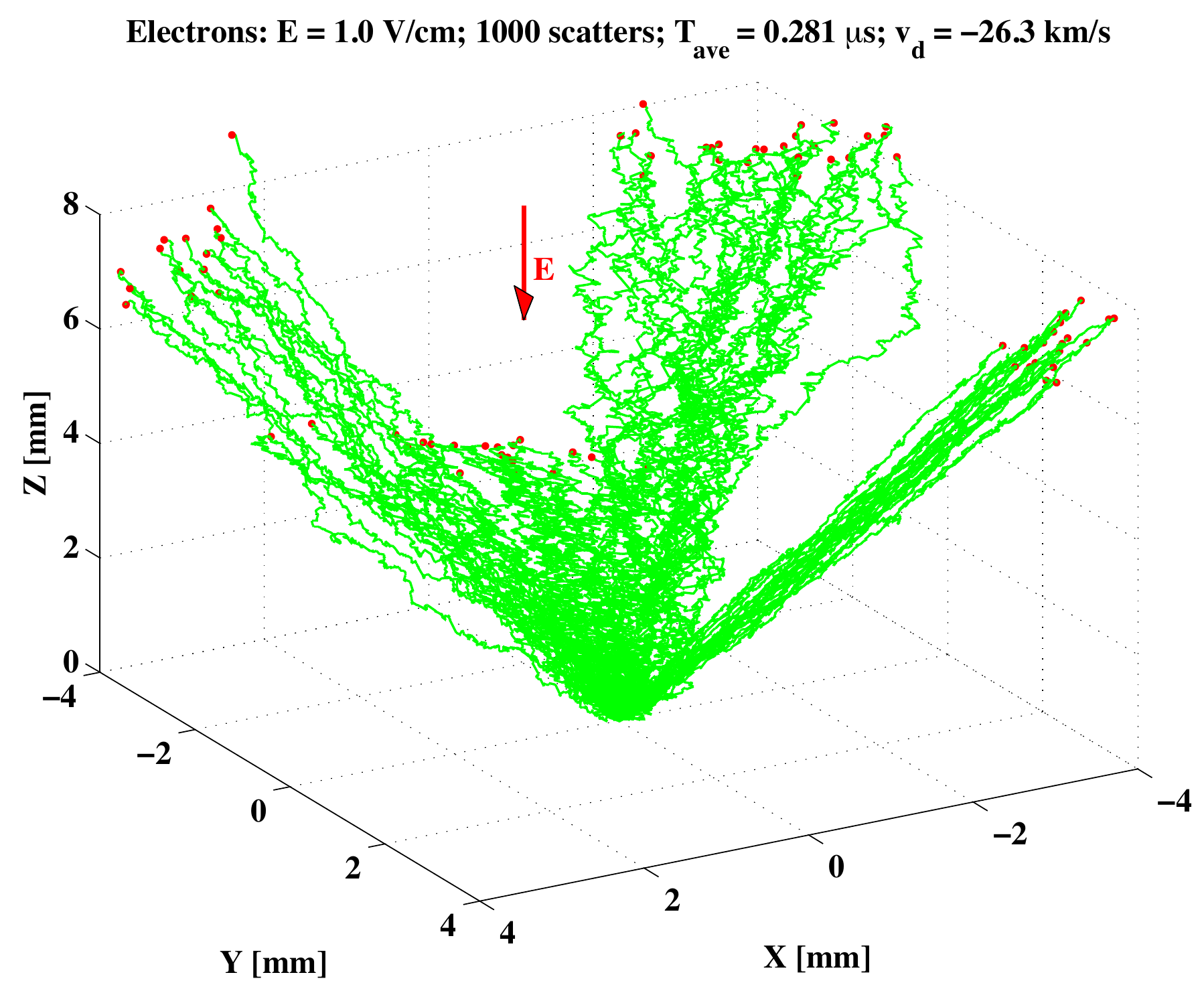} 
  \end{center}
  \caption{ Simulation of electron propagation through Ge cyrstal at zero temperature with a 1 V/cm applied electric field along the -z axis (x, y, \& z aligned with [100] crystal axes).  Shown are the electron trajectories (green) and the final positions (red).  On average 1000 scatters took 0.28 $\mu$s, traversed 7.4 mm along the z-axis. There is large anisotropy with four elliptical cross section beams with the plane of each near perpendicular to one [111] axis (each beam is 33.2 deg from z axis). The ratio of elliptical spreading is $\sim \sqrt{m_\parallel /m_\perp}$.}
  \label{fig:electron_sim}
\end{figure}

Next, we model the electrons near the minimum of their anisotropic valleys. When an electric field is applied the rate of change in momentum, $e \vec E = d \vec p / dt = \hbar d \vec k / dt $,  is along the direction of the electric field, but because the effective mass is a highly anisotropic tensor, $\vec p = \bar {\bar m} \vec v$ and $(\bar {\bar m})^{-1} e \vec E = d \vec v / dt$, the rate of change in group velocity is in general far from the electric field direction.  For a coordinate system with the z-axis aligned with the principle axis of the valley and origin at the valley minimum, we have $e  E_i / m_i = d v_i / dt$ and the energy is given by $\epsilon_{\vec k} = \hbar^2 k_1^2 / (2 m_\perp) + \hbar^2 k_2^2 / (2 m_\perp) + \hbar^2 k_3^2 / (2 m_\parallel)$ which are azimuthally symmetric ellipses with principal axes ratio of $\sqrt{m_\parallel / m_\perp} = 4.41$ for Ge.

For each valley, we transform into a momentum space where the equal energy surfaces are spheres with the same energy.~\cite{Herring:1956}  Thus $k_i = \sqrt{m_i / m_c} k_i^*$ and we derive $E_i = \sqrt{m_i / m_c} E_i^*$, $v_i = v_i^* / \sqrt{m_i / m_c} $, $x_i = x_i^* / \sqrt{m_i / m_c} $, where $3/m_c = 1/m_\parallel + 2/m_\perp$, and $\omega = \omega^*$ (since $\epsilon_{\vec k} - \epsilon_{\vec k'} = \epsilon_{\vec k}^* - \epsilon_{\vec k'}^*$).  Note that a uniform field in real space transforms to a uniform field in the starred space but magnitude and direction change.  The only remaining issue is the speed of sound, which if isotropic in real space is anisotropic in the starred space.  We assume the speed of sound is isotropic in starred space, using the value from real space.  Within this framework, we perform the entire simulation for electron propagation in starred space with starred electric fields, then we transform back into real space rotating the z-axis back to the four [111] valleys.  The simulation equations are identical to Eqs~\ref{equ:1}-\ref{equ:3} where all parameters are replaced with their starred partners.

A typical result is shown in Fig.~\ref{fig:electron_sim} where we have populated the four Ge valleys with 80 electrons total and follow their trajectories in real space with the electric field along the [100] crystal axis (-z axis).  After the first few mm's, the electrons in the four momentum space valleys become separated in real space.  The spread for each valley around each of the four drift velocity directions has a similar average value to the isotropic hole spread, but the shape is elliptical with the ratio $\sim \sqrt{m_\parallel / m_\perp}$.  As shown in Fig.~\ref{fig:drift_compare}, the simulations are compared with the drift velocity measurements for electrons~\cite{Sundqvist:2009} and fit the data with $\Xi_{\text fit} = 11.0$ eV. Again the shape of the electron drift velocity versus electric field is reproduced.

Oblique propagation of electrons is absent in [100] Si crystals because the six Si valleys all have one principle axes aligned with the electric field, but would appear for [111] Si crystals. For Ge these effects are reduced in [111] crystals, but remain because only one of the valleys has principle axes aligned with the electric field.

In conclusion, we now understand the large spreading in the propagation of electrons versus holes in Ge crystals at temperatures below $\sim$1~K and electric fields below $\sim$5 V/cm as due to the freeze out of inter-valley scattering in the highly anisotropic electron bands.~\cite{Cabrera:2010}  Note that when transitions between valleys turn on, the sum of all valleys becomes isotropic.     We thank A. Broniatowski for valuable discussions on oblique propagation.  This research was funded in part by the Department of Energy (Grant Nos. DE-FG02-04ER41295 and DE- FG02-07ER41480) and by the National Science Foundation (Grant Nos. PHY-0542066, PHY-0503729, PHY-0503629, PHY-0504224, PHY-0705078, PHY-0801712).


\begin{thebibliography}{9}

\bibitem{Akerib:2005}
D. S. Akerib, {et al.}, \emph{Phys. Rev.} {\bf D72}, 052009 (2005).

\bibitem{Luke:1988}
P.N.~Luke, \emph{J. Appl. Phys.} {\bf 64}, 6858 (1988).

\bibitem{Cabrera:2010}
B.~Cabrera, {et al.}, in preparation for \emph{Appl. Phys. Lett.}.

\bibitem{Jacoboni:1983}
C.~Jacoboni, L.~Reggiani, \emph{Rev. Mod. Phys.} {\bf 55}, 645 (1983)

\bibitem{Ridley:1982}
See for example: R.K.~Ridley, {\bf Quantum Processes in Semiconductors}  (Oxford Press, 1982), and references therein.

\bibitem{Herring:1956}
C.~Herring and E.~Vogt, \emph{Phys. Rev.} {\bf 101} 944 (1956).

\bibitem{Sasaki:1958}
W.~Sasaki, {et al.},  \emph{J. Phys. Soc. Japan} {\bf 13} 456 (1958).

\bibitem{Sundqvist:2009}
K.M.~Sundqvist, \emph{AIP Conference Series} {\bf 1185} 128 (2009).

\bibitem{Aubry:2009}
V.~Aubry-Fortuna, \emph{AIP Conference Series} {\bf 1185} 639 (2009).

\end{thebibliography}
\end{document}